\documentstyle[prd,aps,preprint,tighten,epsfig]{revtex}

\begin{document}

\draft

\title{Neutrino Mixing and Leptogenesis in Type II Seesaw Mechanism }
\author{{\bf Wan-lei Guo}\footnote{E-mail: \text{guowl@mail.ihep.ac.cn}} }
\address{CCAST (World Laboratory), P.O. Box 8730,
Beijing 100039, China \\
Institute of High Energy Physics, Chinese Academy of Sciences, \\
P.O. Box 918 (4), Beijing 100039, China\footnote{Mailing
address}\\ Graduate School of the Chinese Academy of Sciences}

\maketitle

\begin{abstract}
In the framework of type II seesaw mechanism we propose two simple
but instructive ans$\rm\ddot{a}$tze for neutrino mixing and
leptogenesis. In each ansatz, the effective Majorana neutrino mass
matrix is composed of two parts --- the part with ${\rm Z_2}$
symmetry arises from the ordinary type I seesaw mechanism and the
part with ${\rm S_3}$ symmetry arises from an additional Higgs
triplet vacuum expectation value. The two ans$\rm\ddot{a}$tze can
simultaneously account for the current neutrino oscillation data
and the cosmological baryon number asymmetry via leptogenesis.
\end{abstract}

\pacs{PACS number(s): 14.60.Pq, 13.10.+q, 25.30.Pt}

\section{Introduction}

In the minimal standard model of electroweak interactions, the
lepton number conservation is assumed and neutrinos are exactly
massless Weyl particles. However, recent Super-Kamiokande
\cite{SK}, SNO \cite{SNO}, KamLAND \cite{KM} and K2K \cite{K2K}
neutrino oscillation experiments have provided us with very strong
evidence that neutrinos are actually massive and lepton flavor
mixing does exist. The ordinary (Type I) seesaw mechanism
\cite{SEESAW} gives a very simple and appealing explanation of the
smallness of left-handed neutrino masses -- it is attributed to
the largeness of right-handed neutrino masses. Furthermore, the
lepton-number-violating and out-of-equilibrium decays of heavy
right-handed neutrinos may lead to the cosmological baryon
asymmetry via leptogenesis \cite{LEP}. The latter makes the seesaw
mechanism more attractive. On the other hand, if there is an
additional ${\rm SU(2)_{\rm L}}$ Higgs triplet, it may also
contribute to the neutrino masses and the cosmological baryon
asymmetry. In this scenario, the Lagrangian relevant for neutrino
masses reads:
\begin{eqnarray}
-{\cal L} & = & \frac{1}{2} \overline{N^c_{\rm R}}  M_{\rm R}
N_{\rm R} + M^2_\Delta {\rm Tr}(\Delta_{\rm L}^\dagger \Delta_{\rm
L})
+ \overline{\psi_{\rm L}}\; Y_\nu N_{\rm R} H  \nonumber \\
& & + \overline{\psi_{\rm L}^c}\; Y_\Delta i \tau_2 \Delta_{\rm L}
\psi_{\rm L} - \mu H^T i \tau_2 \Delta_{\rm L} H + h.c. \,,
\end{eqnarray}
where $\psi_{\rm L}= (\nu_{\rm L}$, $l_{\rm L})^T$  and
$H=(H^0,H^-)^T$ denote the left-handed lepton doublet and the
Higgs-boson weak isodoublet respectively, $N_{\rm R}$ stands for
the right-handed Majorana neutrino singlets, and
\begin{equation}
\Delta_{\rm L}= \left( \matrix{ \frac{1}{\sqrt{2}}\Delta^+ &
\Delta^{++} \cr \Delta^0 & - \frac{1}{\sqrt{2}} \Delta^+ }\right)
\end{equation}
is the ${\rm SU(2)_L}$ Higgs triplet. After spontaneous gauge
symmetry breaking, one may arrive at the so-called type II seesaw
formula \cite{T2}
\begin{equation}
M_\nu \; = M_{\rm  L} - M_{\rm  D} M^{-1}_{\rm  R} M^T_{\rm  D} \;
= M_\nu ^{II}+ M_\nu ^{I},
\end{equation}
where $M_{\rm D} \equiv Y_\nu \langle H \rangle$ with $\langle H
\rangle = v \simeq 174 {\rm GeV}$ and $M_{\rm  L} \equiv 2
Y_\Delta \langle \Delta^0 \rangle$ with $\langle \Delta^0 \rangle
 \simeq \mu^\ast v^2/ M^2_\Delta$.
$M_\nu ^{I}= - M_{\rm  D} M^{-1}_{\rm  R} M^T_{\rm  D}$ is the
ordinary type I seesaw mass term, and the type II seesaw mass term
$M_\nu ^{II}=M_{\rm L}$ arises from the additional Higgs triplet
vacuum expectation value. One advantage for considering the type
II seesaw is that it can naturally explain the degenerate neutrino
masses\cite{AK}.

Recently, some works have been done in the framework of type II
seesaw mechanism. The decay asymmetries of right-handed neutrinos,
including the Higgs triplet contribution, have been calculated in
the standard model and minimal supersymmetric standard model
\cite{Hambye,King,Gu}. The bounds on the lightest right-handed
neutrino mass are also discussed. The deviations from the
bimaximal neutrino mixing can be explained through the type II
seesaw\cite{Rodejohann}. The large atmospheric neutrino mixing
angle requires $b-\tau$ unification in the minimal renormalizable
${\rm SO(10)}$ theory \cite{Bt}.

So far the connections between low-energy neutrino masses and
high-energy leptogenesis have not been detailedly investigated in
the framework of type II seesaw mechanism. In this paper, we
propose two simple but instructive ans$\rm\ddot{a}$tze. They can
reproduce the bi-large neutrino mixing pattern and give rise to
the cosmological baryon asymmetry via leptogenesis. In section II
we first calculate the exact neutrino masses and lepton flavor
mixing angels in the basis where the charged lepton mass matrix is
diagonal, then discuss the corrections of an off-diagonal charged
lepton mass matrix to the lepton flavor mixing matrix. In section
III, we figure out twelve patterns of the Dirac neutrino mass
matrices $M_{\rm D}$ with the help of the type I seesaw formula,
and investigate the leptogenesis in the two ans$\rm\ddot{a}$tze.
Finally the summary and comments are given in section IV.

\section{Neutrino Masses and Mixing}

The light (left-handed) Majorana neutrino mass matrix $M_\nu$ can
be diagonalized by one unitary matrix $U_\nu$ :
\begin{equation}
U_\nu^\dagger M_\nu U_\nu^\ast =  {\rm Diag} \{m_1, m_2, m_3 \}.
\end{equation}
The charged lepton mass matrix $M_l$ is in general non-Hermitian,
hence the diagonalization of $M_l$ needs a bi-unitary
transformation:
\begin{equation}
U_l^\dagger M_l \tilde{U}_l =  {\rm Diag} \{m_e, m_\mu, m_\tau \}.
\end{equation}
The Maki-Nakagawa-Sakata (MNS) lepton flavor mixing
matrix\cite{MNS} $V=U_l^\dagger U_\nu$ can in general be
parametrized as follows:
\begin{equation}
V \; = \; \left ( \matrix{ c_x c_z & s_x c_z & s_z \cr - c_x s_y
s_z - s_x c_y e^{-i\delta} & - s_x s_y s_z + c_x c_y e^{-i\delta}
& s_y c_z \cr - c_x c_y s_z + s_x s_y e^{-i\delta} & - s_x c_y s_z
- c_x s_y e^{-i\delta} & c_y c_z \cr } \right ) . \left ( \matrix{
e^{i\rho} & 0 & 0 \cr 0 & e^{i\sigma} & 0 \cr 0 & 0 & 1 \cr}
\right )\;
\end{equation}
with $s_x \equiv \sin\theta_x$, $c_x \equiv \cos\theta_x$, and so
on. Note that three mixing angles of $V$ can directly be given in
terms of the mixing angles of solar, atmospheric and reactor
\cite{CHOOZ} neutrino oscillations. Namely, $\theta_x \approx
\theta_{\rm  sun}$, $\theta_y \approx \theta_{\rm  atm}$ and
$\theta_z \approx \theta_{\rm  chz}$ hold as a good approximation.
In view of the current experimental data, we have $\theta_x
\approx 33^\circ$ and $\theta_y \approx 46^\circ$ (best-fit values
\cite{Data}) as well as $\theta_z < 12^\circ$. The mass-squared
differences of solar and atmospheric neutrino oscillations are
defined respectively as \cite{Data}
\begin{eqnarray}
 \Delta m^2_{\rm  sun} & \equiv & \; m^2_2 - m^2_1 \;  \approx 6.9 \times 10^{-5} ~ {\rm  eV}^2, \nonumber \\
 \Delta m^2_{\rm  atm} & \equiv & |m^2_3 - m^2_2|   \approx 2.3 \times 10^{-3} ~ {\rm  eV}^2,
\end{eqnarray}

We first assume $M_l$ is diagonal (i.e., $U_l = {\bf 1}$ being a
unity matrix), then consider the corrections of an off-diagonal
charged lepton mass matrix to the lepton flavor mixing matrix. In
this paper, we take $U_\nu$ to has two large mixing angles; namely
$\theta_x \sim 30^\circ$, $\theta_y = 45^\circ$, $\theta_z =
0^\circ$. Without loss of generality, $U_\nu$ can be expressed by
 \begin{equation}
U_\nu \; = \; \left ( \matrix{ c_x  & s_x  & 0 \cr  -
\frac{s_x}{\sqrt{2}}
 &  \frac{c_x}{\sqrt{2}}  & -\frac{\sqrt{2}}{2}  \cr
 -\frac{s_x}{\sqrt{2}}  & \frac{c_x}{\sqrt{2}}
 & \frac{\sqrt{2}}{2}  \cr } \right ) . \left ( \matrix{ e^{i\lambda_1}
& 0 & 0 \cr 0 & e^{i\lambda_2} & 0 \cr 0 & 0 & e^{i\lambda_3} \cr}
\right )\;.
\end{equation}

In the following part of this section, we shall discuss two simple
but instructive ans$\rm\ddot{a}$tze, from which neutrino masses
and lepton flavor mixing angels can be specified when the charged
lepton mass matrix is assumed to be diagonal. Then we discuss the
corrections of an off-diagonal charged lepton mass matrix to the
lepton flavor mixing matrix.

\subsection{Ansatz I}
In ansatz I, the neutrino mass matrix $M_\nu =M_\nu ^{I}+ M_\nu
^{II}$ is given by
\begin{equation}
M_\nu ^{I} \; = \; - \left ( \matrix{ a & 0 & 0 \cr 0 & b & -b \cr
0 & -b & b \cr} \right ), \; \; M_\nu ^{II} \; = \; \left (
\matrix{ a & a & a \cr a & a & a \cr a
& a & a \cr} \right ), \\
\end{equation}
with $a = |a| e^{i \phi}$ and $b = |b| e^{i \psi}$. In Eq.(9), we
have assumed  ${M_\nu^I}_{11}= - {M_\nu^{II}}_{11}$, which implies
${M_\nu}_{11} = 0$. Here $M_\nu ^{I}$ and $M_\nu ^{II}$ (and those
of Ansatz II) have a flavor $2 \leftrightarrow 3$ (${\rm Z_2}$)
symmetry
\footnote{From Eq.(3) we know that ${M_\nu^I}$ and ${M_\nu^{II}}$
are symmetric matrices. In Eqs.(9) and (14), ${M_\nu^I}_{12} =
{M_\nu^I}_{13}$ and ${M_\nu^I}_{22} = {M_\nu^I}_{33}$ can be
determined by $2 \leftrightarrow 3$ (${\rm Z_2}$) symmetry. In
this paper, we have assumed ${M_\nu^I}_{12} = {M_\nu^{I}}_{13} = 0
$ and ${M_\nu^I}_{22} = - {M_\nu^I}_{23}$ in both ansatz I and
ansatz II.}
and a permutation ${\rm S_3}$ symmetry
\footnote{A permutation ${\rm S_3}$ symmetry implies
${M_\nu^{II}}_{11} = {M_\nu^{II}}_{22} = {M_\nu^{II}}_{33}$  and
${M_\nu^{II}}_{12} = {M_\nu^{II}}_{13} = {M_\nu^{II}}_{23}$. For
simplicity, we assume ${M_\nu^{II}}_{11} = {M_\nu^{II}}_{12}$ in
ansatz I and ${M_\nu^{II}}_{11} = 0$ in ansatz II. ${M_\nu^{II}}$
of Eq.(9) (often called the ``democratic" mass matrix ) is a
unique representation of the ${\rm S_3}{\rm (L)} \times {\rm
S_3}{\rm (R)} $ symmetric matrix.}
, respectively. Similar textures of the neutrino mass matrix have
been phenomenologically investigated at low energies\cite{He}. In
Ref.\cite{He}, the same texture as $M_\nu ^{II}$ is introduced as
the perturbation term. However, in this paper, $M_\nu ^{II}$
arises from the additional Higgs triplet vacuum expectation value.
Moreover, our case includes the CP-violating phases.

The neutrino mass matrix $M_\nu$ in Eq.(9) can be diagonalized by
the unitary matrix $U_\nu$ of Eq.(8), with
\begin{equation}
\lambda_1 = \frac{\pi + \phi}{2}, \; \lambda_2 = \frac{ \phi}{2},
\; \lambda_3 = \frac{\pi + \psi}{2}.
\end{equation}
The eigenvalues of $M_\nu$ are
\begin{eqnarray}
 m_1 & = & ( \sqrt{3} - 1 )|a|\;, \nonumber \\
 m_2  & = &  ( \sqrt{3} + 1 )|a|\;, \nonumber \\
 m_3  & = &  2|b|\;.
\end{eqnarray}
The mixing angle $\theta_{x}$ is related to the neutrino masses
$m_1$ and $m_2$:
\begin{equation}
\sin \theta_{x} = \sqrt{\frac{m_1}{m_1 + m_2}} \;.
\end{equation}
With the help of Eqs.(7), (11) and (12) we can derive
\begin{eqnarray}
\theta_{x} & = & \arcsin \sqrt{\frac{3 - \sqrt{3}}{6}} \approx  27.4^\circ, \nonumber \\
 |a| & \approx &  3.2 \times 10^{-3} ~ {\rm  eV}, \nonumber \\
 |b|  & \approx &  2.4 \times 10^{-2} ~ {\rm  eV}.
\end{eqnarray}
It is clear that we have known all the mixing angels and neutrino
masses ($m_1 \approx 2.3 \times 10^{-3} ~ {\rm  eV}, ~ m_2 \approx
8.7 \times 10^{-3} ~ {\rm eV}\; {\rm and} \; m_3 \approx 4.8
\times 10^{-2} ~ {\rm eV}$) in this ansatz, but we don't know the
phases $ \lambda_i $.

\subsection{Ansatz II}
In this case, the neutrino mass matrix $M_\nu = M_\nu ^{I}+ M_\nu
^{II}$ is written as
\begin{equation}
M_\nu ^{I} \; = \; - \left ( \matrix{ a & 0 & 0 \cr 0 & b & -b \cr
0 & -b & b \cr} \right ), \; \; M_\nu ^{II} \; = \; \left (
\matrix{ 0 & d & d \cr d & 0 & d \cr d
& d & 0 \cr} \right ), \\
\end{equation}
where $|a| \ll |b|,\; |d|$, the phases of $a, b$ and $d$ are
defined to be $\phi, \psi$ and $\varphi$ respectively. $M_\nu$ can
also be diagonalized by the unitary matrix $U_\nu$ of Eq.(8). Here
we have neglected $a$ of Eq.(14). The phases $\lambda_i$ can be
expressed as
\begin{equation}
\lambda_1 = \frac{\pi + \varphi}{2}\;, \; \lambda_2 = \frac{
\varphi}{2}\;, \; \lambda_3 = \frac{\pi }{2} + \frac{\arg( 2b + d
) }{2}\;,
\end{equation}
and the eigenvalues of $M_\nu$ are
\begin{eqnarray}
 m_1 & = & |d|\;, \nonumber \\
 m_2  & = &  2 |d|\;, \nonumber \\
 m_3  & = &  |2b + d|\;.
\end{eqnarray}
The mixing angle $\theta_{x}$ is also related to the neutrino
masses $m_1$ and $m_2$ as Eq.(12). Using Eqs.(7), (12) and (16),
we can get
\begin{eqnarray}
\theta_{x} & = & \arcsin \frac{\sqrt{3}}{3}  \approx  35.3^\circ, \nonumber \\
 |d| & \approx &  4.8 \times 10^{-3} ~ {\rm  eV}, \nonumber \\
 |b|  & \approx &  (2.2 \sim 2.7) \times 10^{-2} ~ {\rm  eV}.
\end{eqnarray}
The neutrino masses read $m_1 =  m_2/2 \approx  4.8 \times 10^{-3}
~ {\rm eV}$ and $m_3 \approx  4.9 \times 10^{-2} ~ {\rm  eV} $. At
present, the unitary matrix $U_\nu$ is
\begin{equation}
U_\nu \; = \; \left ( \matrix{ \sqrt{\frac{2}{3}}  &
\frac{1}{\sqrt{3}} & 0 \cr - \frac{1}{\sqrt{6}}
 &  \frac{1}{\sqrt{3}}  & -\frac{\sqrt{2}}{2}  \cr
 -\frac{1}{\sqrt{6}}  & \frac{1}{\sqrt{3}}
 & \frac{\sqrt{2}}{2}  \cr } \right ) . \left ( \matrix{ e^{i\lambda_1}
& 0 & 0 \cr 0 & e^{i\lambda_2} & 0 \cr 0 & 0 & e^{i\lambda_3} \cr}
\right )\; ,
\end{equation}
which implies the tri-bimaximal neutrino mixing. It has been
extensively investigated by several authors\cite{He,Tribi}.

\subsection{Neutrino Mixing Corrected by the Charged Lepton Mass Matrix}

In the above paragraphs, we have assumed $M_l$ to be diagonal and
have thus determined all the neutrino mixing angels: $\theta_{x}
\approx 27.4^\circ (35.3^\circ), \theta_{y} = 45^\circ $ and
$\theta_{z} = 0^\circ$ in ansatz I (ansatz II). Now we consider
the corrections of an off-diagonal charged lepton mass matrix to
the lepton flavor mixing matrix. For simplicity and illustration,
we take $M_l$ to be \cite{Xing0107005}
\begin{equation}
M_l \; =\; \left ( \matrix{ 0  & C_l & 0 \cr C^*_l & B_l  & 0 \cr
0 & 0 & A_l \cr} \right ) \;\; ,
\end{equation}
where $A_l = m_\tau$, $B_l = m_\mu - m_e$, and $C_l =\sqrt{m_e
m_\mu} ~ e^{i\xi}$ with the inputs $m_e = 0.511$ MeV, $m_\mu =
105.658$ MeV, and $m_\tau = 1.777$ GeV \cite{PDG}. Because $M_l$
in Eq.(19) has been assumed to be Hermitian, it can in general be
diagonalized by a unitary matrix
\begin{equation}
U_l \; =\; \left ( \matrix{ \cos\theta    & \sin\theta e^{i\xi} &
0 \cr -\sin\theta e^{-i\xi}  & \cos\theta  & 0 \cr 0   & 0 & 1
\cr} \right ) \;.
\end{equation}
The mixing angle $\theta$ in $U_l$ turns out to be
\begin{equation}
\tan 2\theta \;\; =\;\; 2 ~ \frac{\sqrt{m_e m_\mu}}{m_\mu - m_e}
\;\; .
\end{equation}
Taking into account the hierarchy of charged lepton masses (i.e.,
$m_e \ll m_\mu \ll m_\tau$), one obtains $\sin \theta  \approx
 \sqrt{m_e / m_\mu}$  to a good degree of accuracy.
Using $V=U_l^\dagger U_\nu$, we can derive $\theta_{\rm sun}
\approx 30.1^\circ$, $\theta_{\rm atm} \approx 44.9^\circ$ and
$\theta_{\rm chz} \approx 2.8^\circ$ in ansatz I by assuming $\xi
= \pi$. The value of $\theta_{\rm sun}$ agrees with the current
experimental data in $2 \sigma$ confidence level\cite{Data}. In
ansatz II, $\theta_{\rm sun} \approx 33.1^\circ$, $\theta_{\rm
atm} \approx 44.9^\circ$ and $\theta_{\rm chz} \approx 2.8^\circ$
can be obtained by assuming $\xi = \pi/4$. The result of
$\theta_{\rm sun} \approx 33.1^\circ$ is consistent very well with
the best fit value $\theta_{\rm sun} \approx
33.2^\circ$\cite{Data}. Our numerical analysis show the two
ans$\rm\ddot{a}$tze are compatible with the current experimental
data. The nonvanishing angle $\theta_{\rm chz}$ implies that there
are probably CP-violating effects in neutrino oscillations.


\section{ Leptogenesis }

The decays of heavy Majorana neutrinos, $N_i \rightarrow l +
H^\dagger$ and $N_i \rightarrow l^c + H$, violate both the lepton
number conservation and the CP symmetry. A CP-violating asymmetry
$\varepsilon_i^N$ can be generated through the interference
between the tree-level and one-loop decay amplitudes. In the type
II seesaw mechanism, because of the presence of a Higgs triplet,
there are other two decay processes $N_i \rightarrow l +
H^\dagger$ (contribution from exchanging a virtual Higgs triplet)
and $\Delta_{\rm L} \rightarrow l + l$, which can generate the CP
asymmetries $\varepsilon_i^\Delta$ and $\varepsilon_\Delta$,
respectively. For simplicity, in the following parts of this
paper, we consider the case that three heavy Majorana neutrinos
$N_i$ have a hierarchical mass spectrum ($ M_1 \ll M_2 \ll M_3 $)
and the triplet mass $M_\Delta$ is much larger than $M_1$ ($M_1
\ll M_\Delta$). Then the interactions of $N_i$ can be in thermal
equilibrium when $N_2$, $N_3$ and $\Delta_{\rm L}$ decay. The
CP-violating asymmetries produced in the decays of $N_2$, $N_3$
and $\Delta_{\rm L}$ can be erased before $N_i$ decays. Then only
the asymmetries $\varepsilon_1^N$ and $\varepsilon_1^\Delta$
produced by out-of-equilibrium decay of $N_1$ survives. In a
flavor diagonal basis for the heavy right-handed neutrinos one has
\cite{Hambye,King,Gu}
\footnote{In this note, we take the result of Ref.\cite{King} for
$\varepsilon_1^\Delta $.}
\begin{eqnarray}
\varepsilon_1^N \approx - \frac{3}{16 \pi v^2} \;
\frac{M_1}{(M_{\rm   D}^\dagger M_{\rm   D})_{11}} \; {\rm
Im}[(M_{\rm   D}^T (m_\nu^{I})^* M_{\rm   D})_{11}] \; ,
\end{eqnarray}
and
\begin{eqnarray}
\varepsilon_1^\Delta \approx - \frac{3}{16 \pi v^2} \;
\frac{M_1}{(M_{\rm   D}^\dagger M_{\rm D})_{11}} \; {\rm
Im}[(M_{\rm   D}^T (m_\nu^{II})^* M_{\rm   D})_{11}] \; .
\end{eqnarray}
The $\varepsilon_1^N $ and $\varepsilon_1^\Delta $ can result in a
net lepton number asymmetry
\begin{eqnarray}
Y_{\rm  L} \equiv \frac{n_{\rm  L} - n_{\bar{{\rm  L}}}}{\bf s} =
\frac{\kappa}{g_*} \; \varepsilon_1 ,
\end{eqnarray}
where $\varepsilon_1 = \varepsilon_1^N + \varepsilon_1^\Delta$,
$g_* = 106.75$ is an effective number characterizing the
relativistic degrees of freedom which contribute to the entropy
{\bf s} of the early universe, and $\kappa$ accounts for the
dilution effects induced by the lepton-number-violating wash-out
processes. The dilution factor $\kappa$ can be determined by
solving the full Boltzmann equations\cite{Review}. Because the
Higgs triplet $\Delta_{\rm L}$ does not couple directly to the
lightest right-handed neutrino $N_1$, the scatterings with a
virtual triplet can not change the number density of $N_1$ in the
Boltzmann equations. On the other hand, the additional scattering
$l + l \rightarrow \Delta_{\rm L} \rightarrow H + H$ is largely
suppressed \cite{Hambye}, so we neglect the contribution to
dilution effects from the scatterings involving the additional
Higgs triplet. We take the ordinary dilution factor formulas of
type I seesaw\cite{Kappa}:
\begin{eqnarray}
\kappa \; & = & \; 0.3 \left (\frac{10^{-3} ~ {\rm
eV}}{\tilde{m}_1} \right ) \left [ \ln \left (
\frac{\tilde{m}_1}{10^{-3} ~ {\rm  eV}} \right ) \right ]^{-0.6}
\;, \; 10^{-2} ~ {\rm  eV} \lesssim \tilde{m}_1 \lesssim 10^3 ~
{\rm  eV} \cr \kappa \; & = & \frac{1}{\displaystyle 2 \sqrt{\left
( \frac{\tilde{m}_1}{10^{-3} ~ {\rm  eV}} \right )^2+9}} \; , \;
\; ~   \tilde{m}_1 \lesssim 10^{-2} ~ {\rm eV}
\end{eqnarray}
with $\tilde{m}_1 = ({M_{\rm D}}^{\dag}\,M_{\rm D})_{11} /
{M_1}$\cite{BP}. The lepton number asymmetry $Y_{\rm  L}$ is
eventually converted into a net baryon number asymmetry $Y_{\rm
B}$ via the nonperturbative sphaleron processes \cite{Kuzmin}:
$Y_{\rm  B} \approx -0.55 Y_{\rm  L}$. A generous range $0.7
\times 10^{-10} \lesssim Y_{\rm  B} \lesssim 1.0 \times 10^{-10}$
has been drawn from the recent WMAP observational data
\cite{WMAP}.

\subsection{Determining the Structure of $M_{\rm D}$ }
From Eqs.(9) and (14) we know
\begin{equation}
M_\nu^I \; =  - M_{\rm  D} M^{-1}_{\rm  R} M^T_{\rm  D} \; = -
\left ( \matrix{ a & 0 & 0 \cr 0 & b & -b \cr 0 & -b & b \cr}
\right )\;,
\end{equation}
where $M_{\rm  R}$ is assumed to be diagonal. It is worth
mentioning that $M_\nu^I$ has the two-zero texture\cite{Twozero}.
The following conditions can be derived from Eq.(26):
\begin{eqnarray}
{M_\nu^I}_{12} & = & {M_\nu^I}_{13} = 0, \nonumber \\
{M_\nu^I}_{22}  & = &  {M_\nu^I}_{33} = - {M_\nu^I}_{23} \neq 0
\;.
\end{eqnarray}
In terms of the above conditions, we can obtain some constraints
on $M_{\rm D}$. In order to fix the structure of $M_{\rm D}$, we
exclude the possible cancellations among the elements of $M_{\rm
D}$ and $M_{\rm R}$, and require that the above conditions should
be from the zeros and relations of the elements in $M_{\rm D}$. We
find that there are twelve patterns of $M_{\rm  D}$ which can
produce the texture of $M_\nu^I$ through the type I seesaw
formula. We have listed all of them in Table 1 and have classified
them into four categories according to their structures.

\subsection{Leptogenesis for Ansatz I}

We find that only patterns A1 and A2 in Table 1 can generate
nonvanishing $\varepsilon_1$. It is worth remarking that patterns
C1 and D1 give rise to $\varepsilon_1^N = - \varepsilon_1^\Delta
\neq 0 $, which means $\varepsilon_1 = 0$. Patterns C3 and D3 lead
to $(M_{\rm D}^\dagger M_{\rm D})_{11} = 0$ for the two given
ans$\rm\ddot{a}$tze, implying that $N_1$ does not decay. So we
neglect these two textures in this paper. Because A1 and A2 have
similar physical consequences, we shall only focus on pattern A1.

In pattern A1, the Dirac neutrino mass matrix $M_{\rm D}$ is
\begin{eqnarray}
M_{\rm   D} = \left ( \matrix{ 0 & 0 & C \cr A & B & 0 \cr -A & -B
& 0 \cr} \right ) \;.
\end{eqnarray}
It is then straightforward to obtain  $\tilde{m}_1 = 2 \left| A^2
\right|/M_1$ . With the help of Eqs.(22), (23) and (26), we have
\begin{eqnarray}
\varepsilon_1^N & = & - \frac{3}{8 \pi v^2} M_1 \left|
\frac{B^2}{M_2} \right| \sin 2 (\beta - \alpha) \; , \cr
\varepsilon_1^\Delta & = & 0 \; .
\end{eqnarray}
Using Eqs.(13) and (26), we can derive
\begin{eqnarray}
 |a| & = &  \left | \frac{C^2}{M_3} \right | \approx 3.2 \times
10^{-3} ~ {\rm  eV } \;, \cr |b| & = & \left | \frac{A^2}{M_1} +
\frac{B^2}{M_2} \right | \approx 2.4 \times 10^{-2} ~ {\rm  eV }
\; .
\end{eqnarray}
The relations between $\left | A^2 \right | / M_1 $ and $\left |
B^2 \right | / M_2 $ shown in Eq.(30) are very important, because
they affect the dilution factor $\kappa$ (i.e., $\tilde{m}_1$) and
$\varepsilon_1$. For illustration, we discuss two special cases:

(1) In the  $\left | A^2 \right | / M_1 \gg \left | B^2 \right | /
M_2$ case, $\left | A^2 \right | / M_1 \approx 2.4 \times 10^{-2}
~ {\rm  eV }$ and $\left | B^2 \right | / M_2 \ll 2.4 \times
10^{-2} ~ {\rm  eV }$. Therefore we use the first formula of
Eq.(25) to calculate the dilution factor. With the help of
Eqs.(24), (25) and (30), we have
\begin{eqnarray}
Y_{\rm  B} \approx -0.55 \frac{\kappa}{g_*} \; \varepsilon_1
\approx 1.7 \times 10^{-6} \frac{M_1}{v^2} \left| \frac{B^2}{M_2}
\right| \sin 2 (\beta - \alpha)\; .
\end{eqnarray}
For example, when we take $\left | B^2 \right | / M_2 = 10^{-3} \;
{\rm eV}$, the successful leptogenesis requires $1.2 \times
10^{12} \; {\rm GeV} \lesssim M_1 \lesssim 1.8 \times 10^{12} \;
{\rm GeV}$ by assuming $\sin 2 (\beta - \alpha) =1$.

(2) In the  $\left | A^2 \right | / M_1 \ll \left | B^2 \right | /
M_2$ case, $\left | B^2 \right | / M_2 \approx 2.4 \times 10^{-2}
~ {\rm  eV }$ and $\left | A^2 \right | / M_1 \ll 2.4 \times
10^{-2} ~
 {\rm  eV }$. Using the second formula
of Eq.(25) we find  $0.048 < \kappa \leq 0.167 $. In this case the
baryon asymmetry $Y_{\rm B}$ is
\begin{eqnarray}
Y_{\rm  B} \approx  8.5 \times 10^{-17} \; \kappa  \frac{M_1}{v}
\sin 2 (\beta - \alpha)\;.
\end{eqnarray}
Here we take  $\sin 2 (\beta - \alpha) = 1$, $M_1 = 10^9 \; { \rm
GeV}$ and $\kappa = 0.16$, only for illustration. The baryon
asymmetry is found to be $Y_{\rm B} \approx 0.78 \times 10^{-10}$,
which is compatible with the recent WMAP observational data. So
this case can produce the cosmological baryon asymmetry via
leptogenesis.

\subsection{Leptogenesis for Ansatz II}

Through a global analysis we find that seven patterns (A1, A2, B1,
B2, B3, C1 and D1) in Table 1 can generate the cosmological baryon
asymmetry via leptogenesis. It is worth remarking that A1 and A2
(B1, B2 and B3; C1 and D1) have similar physical consequences,
therefore we concentrate on patterns A1, B1 and C1 as three
typical examples for numerical illustration.

For pattern A1, $\tilde{m}_1 = 2 \left | A^2 \right | / M_1$ and
\begin{eqnarray}
\varepsilon_1^N & = & - \frac{3}{8 \pi v^2} M_1 \left| \frac{
B^2}{M_2} \right| \sin 2 ( \beta - \alpha ) \; , \cr
\varepsilon_1^\Delta & = & - \frac{3}{16 \pi v^2} M_1 |{d}| \sin (
\varphi - 2\alpha )\; ,
\end{eqnarray}
with
\begin{eqnarray}
 |a| & = & \left | \frac{{\rm  C^2}}{M_3} \right | \; \ll |d| \approx 4.8 \times
10^{-3} ~ {\rm  eV } \;, \cr |b| & = & \left | \frac{ A^2}{M_1} +
\frac{ B^2}{M_2} \right | \approx 2.5 \times 10^{-2} ~ {\rm  eV }
\;.
\end{eqnarray}
For simplicity, we have taken the average values of $|b|$ in
Eq.(17). If $\sin ( \varphi - 2\alpha ) = 0$ or $|d| \sin (
\varphi - 2\alpha ) \ll \left | B^2 \right | / M_2 \; \sin 2 (
\alpha - \beta )$ (i.e., $\varepsilon_1^N \gg
\varepsilon_1^\Delta$ ) holds, we shall obtain similar results as
shown in pattern A1 of ansatz I. If we assume $\sin 2 ( \beta -
\alpha ) = \sin ( \varphi - 2\alpha ) = 1$ and then consider the
$\left | d \right | \gg \left | B^2 \right | / M_2$ case (thus we
neglect the contribution of $\varepsilon_1^N$), the baryon
asymmetry is found to be $ Y_{\rm B} \approx  2.2 \times 10^{-20}
\; M_1 /v $. It is clear that the successful leptogenesis requires
$ 5.5 \times 10^{11}\; { \rm GeV} \lesssim M_1 \lesssim 7.9 \times
10^{11}\; { \rm GeV}$.

Now let us focus on pattern B1
\begin{eqnarray}
M_{\rm   D} = \left ( \matrix{ 0 & B & C \cr A & 0 & 0 \cr -A & 0
& 0 \cr} \right ) \; .
\end{eqnarray}
We have $\tilde{m}_1 =2 \left | A^2 \right | / M_1$, and
\begin{eqnarray}
\varepsilon_1^N & = & 0 \; , \cr \varepsilon_1^\Delta & = & -
\frac{3}{16 \pi v^2} M_1 |{d}| \sin (\varphi - 2 \alpha)\;  ,
\end{eqnarray}
with
\begin{eqnarray}
 |a| & = & \left |\frac{ B^2}{M_2} + \frac{C^2}{M_3} \right | \; \ll |d| \approx 4.8 \times
10^{-3} ~ {\rm  eV } \;, \cr |b| & = & \left | \frac{ A^2}{M_1}
\right | \approx 2.5 \times 10^{-2} ~ {\rm  eV } \; .
\end{eqnarray}
In this pattern we find the baryon asymmetry $Y_{\rm B} \approx
2.2 \times 10^{-20} \;  M_1  /v \; \sin (\varphi - 2 \alpha) $.
The recent observational data of $Y_{\rm B}$ requires $M_1 \geq
5.5 \times 10^{11} \; {\rm GeV}$.

In pattern C1, the Dirac neutrino mass matrix $M_{\rm D}$ is
\begin{eqnarray}
M_{\rm D} = \left ( \matrix{ B & 0 & C \cr 0 & A & 0 \cr 0 & -A &
0 \cr} \right ) \; .
\end{eqnarray}
In this pattern $\tilde{m}_1 = \left | B^2 \right | / M_1$. The CP
asymmetries $\varepsilon_1^N$ and $\varepsilon_1^\Delta$ are
\begin{eqnarray}
\varepsilon_1^N & = & - \frac{3}{16 \pi v^2} M_1 \left| \frac{
C^2}{M_3} \right| \sin 2 (\gamma - \beta) \;, \cr
\varepsilon_1^\Delta & = & 0 \; .
\end{eqnarray}
The constraints from Eqs.(17) and (26) are
\begin{eqnarray}
 |a| & = &  \left | \frac{B^2}{M_1} +  \frac{ C^2}{M_3} \right | \ll |d| \approx 4.8 \times
10^{-3} ~ {\rm  eV } \;, \cr |b| & = & \left |\frac{ A^2}{M_2}
\right | \approx 2.5 \times 10^{-2} ~ {\rm  eV } \; .
\end{eqnarray}
For illustration, here we assume $ \left | B^2 \right | / M_1 =
\left | C^2 \right | / M_3 = 10^{-4} \; {\rm  eV}$ and $\sin 2
(\gamma - \beta) = 1$. This produces a baryon asymmetry $Y_{\rm B}
\approx 2.9 \times 10^{-20} \;  M_1/v \; $, which in turn requires
$4.2 \times 10^{11}\; { \rm GeV} \lesssim M_1 \lesssim 6.0 \times
10^{11}\; { \rm  GeV}$.

Because the three light neutrinos have a clear mass hierarchy in
ansatz I and ansatz II, we have neglected possible
renormalization-group running effects of neutrino masses and
lepton flavor mixing parameters between the scales $v$ and $M_1$
\cite{RGE}.

Finally, let us comment on the relation between leptogenesis and
the CP violation at low energies. Because of the corrections from
the off-diagonal charged lepton mass matrix $M_l$, we do not find
a direct link between $\varepsilon_1$ and the low-energy CP
violation in neutrino oscillations, but we can derive some
relations among the CP-violating phases. For example, we have
$\psi = 2 \alpha$ in pattern  B1 of ansatz II and $\psi = \arg
(|A|^2 / M_1\; e^{2 i \alpha}  + |B|^2 / M_2\; e^{2 i \beta})$ in
pattern A1 of ansatz I, and so on. If the charged lepton mass
matrix $M_l$ is real (i.e., $\xi = 0$), then there is no CP
violation in neutrino oscillations, because the Dirac CP-violating
phase $\delta = 0$ holds.

\section{Summary and Comments}

We have proposed two simple but instructive ans$\rm\ddot{a}$tze of
neutrino mass matrix in type II seesaw mechanism. The explicit
neutrino masses and bi-large neutrino mixing can be calculated
when the charged lepton mass matrix is assumed to be diagonal. We
have also discussed the possible corrections of an off-diagonal
charged lepton mass matrix to the lepton flavor mixing matrix, and
the results are compatible with the current neutrino oscillation
data. On the other hand, we have figured out twelve patterns of
the Dirac neutrino mass matrix $M_{\rm D}$ with the help of the
type I seesaw formula, and have investigated the leptogenesis in
the two ans$\rm\ddot{a}$tze. Two patterns of $M_{\rm D}$ in ansatz
I and seven patterns of $M_{\rm D}$ in ansatz II can generate the
desired cosmological baryon asymmetry. We have briefly discussed
the relation between leptogenesis and the CP violation at low
energies.

It is worth mentioning that one can carry out a similar analysis
of $Y_{\rm B}$ in the framework of supersymmetric type II seesaw
and leptogenesis models.  If the future KATRIN\cite{Ka} and WMAP
experiments can pin down the absolute neutrino masses\cite{Xing},
this will be very helpful to examine the two ans$\rm\ddot{a}$tze.
Our results will be very useful for model building, in order to
understand why neutrino masses are so tiny and why two of the
lepton flavor mixing angles are so large.

\acknowledgments{I would like to thank my supervisor, Z.Z. Xing,
for stimulating discussions and reading the manuscript. I am also
grateful to P.H. Gu and X.J. Bi for helpful communication. This
work was supported in part by the National Nature Science
Foundation of China.}


\newpage
\begin{table}
\caption{Twelve patterns of the Dirac neutrino mass matrix $M_{\rm
D}$ derived from Eq.(26). They may be favored or disfavored by the
current cosmological baryon asymmetry $Y_{\rm  B}$. In this paper,
$A, B$ and $C$ are complex, and their phases are $\alpha$, $\beta$
and $\gamma$ respectively.}
\begin{center}
\begin{tabular}{cccc}   \\
Pattern & By current data & Pattern & By current data   \\ \\ \hline \\
$\rm  A_1: \left ( \matrix{ 0 & 0 & C \cr A & B & 0 \cr -A & -B &
0 \cr} \right )$
 & $\matrix {\hspace{-0.4cm}{\rm Ansatz \;\; I \; \; favored} \cr \hspace{-0.4cm}{\rm  Ansatz \; II \; \; favored} }$
 & $\rm  C_1: \left ( \matrix{ B & 0 & C \cr 0 & A &
0 \cr 0 & -A & 0 \cr} \right )$
 & $\matrix {{\rm  Ansatz \;\; I \; \; disfavored} \cr \hspace{-0.4cm} {\rm  Ansatz \; II \; \; favored} }$
\\ \\ \hline\\
$\rm  A_2: \left ( \matrix{ 0 & C & 0 \cr A & 0 & B \cr -A & 0 &
 -B \cr} \right )$
 & $\matrix {\hspace{-0.4cm}{\rm  Ansatz \;\; I \; \; favored} \cr\hspace{-0.4cm} {\rm  Ansatz \; II \; \; favored} }$
 & $\rm  C_2: \left ( \matrix{ B & 0 & 0 \cr 0 & A & 0 \cr
0 & -A & 0 \cr} \right )$
 & $\matrix {{\rm  Ansatz \;\; I \; \; disfavored} \cr {\rm  Ansatz \; II \; \; disfavored} }$
\\ \\ \hline\\
$\rm  A_3: \left ( \matrix{ C & 0 & 0 \cr 0 & A & B \cr 0 & -A &
-B \cr} \right )$
 & $\matrix {{\rm  Ansatz \;\; I \; \; disfavored} \cr {\rm  Ansatz \; II \; \; disfavored} }$
 & $\rm  C_3: \left ( \matrix{ 0 & 0 &
 C \cr 0 & A & 0 \cr 0 & -A & 0 \cr} \right )$
 & $\matrix {{\rm  Ansatz \;\; I \; \; disfavored} \cr {\rm  Ansatz \; II \; \; disfavored}}$
\\ \\ \hline\\
$\rm  B_1: \left ( \matrix{ 0 & B & C \cr A & 0 & 0 \cr -A & 0 & 0
\cr} \right )$
 & $\matrix {{\rm  Ansatz \;\; I \; \; disfavored} \cr \hspace{-0.45cm} {\rm  Ansatz \; II \; \; favored} }$
 & $\rm  D_1: \left ( \matrix{ B & C & 0 \cr 0 & 0 & A \cr 0 & 0 & -A \cr} \right)$
 & $\matrix {{\rm  Ansatz \;\; I \; \; disfavored} \cr \hspace{-0.4cm} {\rm  Ansatz \; II \; \; favored} }$
\\ \\ \hline\\
$\rm  B_2: \left ( \matrix{ 0 & B & 0 \cr A & 0 & 0 \cr -A & 0 & 0
\cr} \right )$
 & $\matrix {{\rm  Ansatz \;\; I \; \; disfavored} \cr \hspace{-0.45cm} {\rm  Ansatz \; II \; \; favored} }$
 & $\rm  D_2: \left ( \matrix{ B & 0
& 0 \cr 0 & 0 & A \cr 0 & 0 & -A \cr} \right )$
 & $\matrix {{\rm  Ansatz \;\; I \; \; disfavored} \cr {\rm  Ansatz \; II \; \; disfavored} }$
\\ \\ \hline\\
$\rm  B_3: \left ( \matrix{ 0 & 0 & C \cr A & 0 & 0 \cr -A & 0 & 0
\cr} \right )$
 & $\matrix {{\rm  Ansatz \;\; I \; \; disfavored} \cr \hspace{-0.45cm} {\rm  Ansatz \; II \; \; favored}}$
 & $\rm  D_3: \left ( \matrix{ 0 & C
& 0 \cr 0 & 0 & A \cr 0 & 0 & -A \cr} \right )$
 & $\matrix {{\rm  Ansatz \;\; I \; \; disfavored} \cr {\rm  Ansatz \; II \; \; disfavored} }$
\\  \\
\end{tabular}
\end{center}
\end{table}
\end{document}